\documentclass[useAMS,usenatbib]{mnras}
\usepackage[dvips]{graphicx}
\usepackage[english]{babel}
\usepackage{amsmath}
\usepackage{amssymb}
\usepackage{textcomp}
\usepackage{natbib}
\usepackage{wasysym}

\title[The deuterium abundance and stellar mass loss]{On the deuterium abundance and the importance of stellar mass loss in the interstellar and intergalactic medium} 
\author[F. van de Voort et al.]{Freeke~van~de~Voort,$^{1,2}$\thanks{E-mail: freeke.vandevoort@h-its.org}
Eliot~Quataert,$^{3}$
Claude-Andr\'e~Faucher-Gigu\`ere,$^{4}$
\newauthor
Du\v{s}an~Kere\v{s},$^{5}$
Philip~F.~Hopkins,$^{6}$ 
T.~K.~Chan,$^5$
Robert~Feldmann$^{7}$ and
\newauthor
Zachary~Hafen$^4$ \\
$^{1}$Heidelberg Institute for Theoretical Studies, Schloss-Wolfsbrunnenweg 35, 69118, Heidelberg, Germany \\
$^{2}$Astronomy Department, Yale University, PO Box 208101, New Haven, CT 06520-8101, USA \\
$^{3}$Department of Astronomy and Theoretical Astrophysics Center, University of California, Berkeley, CA 94720-3411, USA \\
$^{4}$Department of Physics and Astronomy and CIERA, Northwestern University, 2145 Sheridan Road, Evanston, IL 60208, USA \\
$^{5}$Department of Physics, Center for Astrophysics and Space Science, University of California at San Diego, 9500 Gilman Drive, \\
\ \ La Jolla, CA 92093 \\
$^{6}$TAPIR, Mailcode 350-17, California Institute of Technology, Pasadena, CA 91125, USA \\
$^{7}$Institute for Computational Science, University of Zurich, Zurich CH-8057, Switzerland 
}

\begin{document}

\date{Accepted 2018 February 28. Received 2018 February 23; in original form 2017 April 26}

\pagerange{\pageref{firstpage}--\pageref{lastpage}} \pubyear{2018}

\maketitle

\label{firstpage}

\begin{abstract}

We quantify the gas-phase abundance of deuterium and fractional contribution of stellar mass loss to the gas in cosmological zoom-in simulations from the Feedback In Realistic Environments project. At low metallicity, our simulations confirm that the deuterium abundance is very close to the primordial value. The chemical evolution of the deuterium abundance that we derive here agrees quantitatively with analytical chemical evolution models. We furthermore find that the relation between the deuterium and oxygen abundance exhibits very little scatter. We compare our simulations to existing high-redshift observations in order to determine a primordial deuterium fraction of $(2.549\pm0.033)\times10^{-5}$ and stress that future observations at higher metallicity can also be used to constrain this value. At fixed metallicity, the deuterium fraction decreases slightly with decreasing redshift, due to the increased importance of mass loss from intermediate-mass stars. We find that the evolution of the average deuterium fraction in a galaxy correlates with its star formation history. Our simulations are consistent with observations of the Milky Way's interstellar medium: the deuterium fraction at the solar circle is $85-92$~per cent of the primordial deuterium fraction. We use our simulations to make predictions for future observations. In particular, the deuterium abundance is lower at smaller galactocentric radii and in higher mass galaxies, showing that stellar mass loss is more important for fuelling star formation in these regimes (and can even dominate). Gas accreting onto galaxies has a deuterium fraction above that of the galaxies' interstellar medium, but below the primordial fraction, because it is a mix of gas accreting from the intergalactic medium and gas previously ejected or stripped from galaxies.

\end{abstract}


\section{Introduction} \label{sec:intro}

Deuterium is one of the few stable isotopes produced in astrophysically interesting amounts during Big Bang nucleosynthesis, together with helium and lithium (see \citealt{Steigman2007} for a review). Helium and lithium can be produced after this initial phase, in stars and via collisions of cosmic ray nuclei, potentially increasing their gas-phase abundances. However, the gas-phase deuterium abundance can only decrease. All primordial deuterium is burned during the collapse of a protostar and deuterium synthesized in stellar interiors is immediately destroyed, because deuterium fuses at relatively low temperatures, $T\approx10^6~K$, easily reached in the interiors of stars and even brown dwarfs \citep{Epstein1976, Stahler1988, Spiegel2011}. Therefore, mass lost from stars (also referred to as `recycled' gas) is deuterium-free, i.e.\ $\mathrm{(D/H)}_\mathrm{recycled}=0$. 

The primordial deuterium fraction, $\mathrm{(D/H)}_\mathrm{prim}$, is sensitive to cosmological parameters and, in particular, to the baryon--photon ratio and thus to the baryonic density of the Universe. Measurements of the cosmic microwave background (CMB) radiation have pinned down the ratio of the mean density of baryons to the critical density and the Hubble parameter \citep{Planck2015XIII}. The most recent theoretical models of Big Bang nucleosynthesis have incorporated these and derived, for example, $\mathrm{(D/H)}_\mathrm{prim}=(2.45\pm0.05)\times10^{-5}$ \citep{Coc2015} and $\mathrm{(D/H)}_\mathrm{prim}=(2.58\pm0.13)\times10^{-5}$ \citep{Cyburt2016}, where the quoted errors are $1\sigma$. 

An accurate determination of the primordial deuterium fraction, in conjunction with Big Bang nucleosynthesis reaction rates, gives an independent constraint on the cosmic baryon density. If there is disagreement with the value derived from CMB measurements, this could point to a deviation in the expansion rate of the early universe and to non-standard models of big bang nucleosynthesis. Low-metallicity gas likely has a deuterium fraction close to the primordial value, because it has not been substantially enriched by stellar mass loss. Absorption lines in spectra of background quasars have been used to determine the primordial deuterium fraction observationally, finding e.g.\ $\mathrm{(D/H)}_\mathrm{prim}=(2.547\pm0.033)\times10^{-5}$ \citep{Cooke2016}. Modern estimates are thus consistent with each other and there is currently no conflict with the standard model of cosmology \citep[e.g.][]{Steigman2007, Cooke2014}. 

Intermediate-mass and massive stars return material to the interstellar medium (ISM) via stellar winds before and during the asymptotic giant branch (AGB) phase and via supernova explosions, respectively. One well-known effect of this recycling process of baryons that become part of a star and are later returned into space is the release of metals into the ISM and the intergalactic medium (IGM). However, it is also important for the destruction of light elements, such as deuterium. If there is no fresh infall of gas onto galaxies and the ISM of these objects is replenished by stellar mass loss, both the metallicity of the gas and young stars increases and the deuterium fraction in the ISM decreases. The ratio of the deuterium fraction in the ISM or IGM and the primordial deuterium fraction, $\mathrm{(D/H)}/\mathrm{(D/H)}_\mathrm{prim}$, is therefore a measure of the fraction of the gas that has not been processed in stars. The inverse of this, i.e.\ $\mathrm{(D/H)}_\mathrm{prim}/\mathrm{(D/H)}$, is known as the astration factor. Measurements of the evolution of the deuterium abundance have been used to constrain galactic chemical evolution models \citep[e.g.][]{Audouze1974, Vangioni1988, Vangioni1994, Scully1997, Olive2012}. These models predict astration factors higher than observed in the local ISM when they only take into account cosmological inflow \citep[e.g.][]{Fields1996, Romano2006, Lagarde2012}. Models that additionally allow for galactic outflows predict lower astration factors \citep[e.g.][]{Dvorkin2016, Weinberg2017}. In this way, measurements of the deuterium fraction (and thus the astration factor) can shed light on the balance between primordial inflow, metal-enriched outflow, and recycling through stellar mass loss, which are all related to the star formation and accretion history of a galaxy \citep[e.g.][]{Casse1998, Prantzos2001, Romano2006, Dvorkin2016, Weinberg2017}. 

The fuelling of star formation by stellar mass loss is likely more important in high-mass galaxies and high-density environments. Massive early-type galaxies and satellite galaxies have specific star formation rates (SFRs) far below those of central late-type galaxies. It is not known which process(es) quench(es) galaxies, but galactic outflows are at least partially responsible for quenching massive galaxies and preventing subsequent gas accretion \citep[e.g.][]{Faucher2011, Voort2011a}. However, a substantial fraction of local early-type galaxies still have a detectable molecular or atomic gas reservoir \citep[e.g.][]{Young2011, Serra2012}. 
Some of these exhibit gas kinematics indicating a predominantly external gas supply, such as through minor mergers, whereas others (especially those located in a cluster environment) are consistent with their ISM being fed through stellar mass loss \citep{Davis2011}. Furthermore, some massive galaxies in the centres of clusters are forming stars at a substantial rate ($1-100$~M$_{\astrosun}$~yr$^{-1}$) and contain a considerable amount of dust \citep{ODea2008, Donahue2011}. Dust is produced by stars and destroyed by sputtering in hot gas. Therefore, the gas supply is unlikely to have cooled out of the hot halo gas. This also indicates that stellar mass loss may be an important contributor to the fuel for the observed star formation \citep{Voit2011}. 

The amount of mass supplied to the ISM through stellar mass loss could also be sufficient to fuel most of the star formation in present-day star-forming galaxies, including the Milky Way \citep{Leitner2011}. However, galactic outflows were not included, but are likely required to produce correct stellar masses and metallicities. Hydrodynamical simulations that included feedback from stars and/or black holes have found that stellar mass loss becomes more important for fuelling star formation towards lower redshift, although, in general, it does not become the dominant fuel source for star formation \citep{Oppenheimer2008, Segers2016}. 

The predicted deuterium fraction and the importance of stellar mass loss are the focus of this paper. We present results from a suite of high-resolution, cosmological `zoom-in' simulations from the `Feedback In Realistic Environments' (FIRE) project,\footnote{http://fire.northwestern.edu/} which spans a large range in halo and galaxy mass. The FIRE simulation suite has been shown to successfully reproduce a variety of observations, which is linked to the strong stellar feedback implemented. These galactic winds efficiently redistribute gas from galaxies out to large galactocentric distances (see \citealt{Muratov2015, Muratov2016}). For the purposes of this paper, we highlight the fact that the simulations match the derived stellar--to--halo mass relationship \citep{Hopkins2014FIRE, Feldmann2016}, the galaxy mass-metallicity relation and gas-phase metallicity gradients at $z=0-3$ \citep{Ma2016, Ma2017}, and the dense neutral hydrogen, H\,\textsc{i}, content of galaxy haloes \citep{Faucher2015, Faucher2016, Hafen2017}.  

This is the first time cosmological, hydrodynamical simulations are used to study the deuterium abundance in the ISM and IGM. Our simulations self-consistently follow the time-dependent assembly of dark matter haloes, the accretion of gas onto galaxies, the formation of stars, the return of mass in stellar winds, and the generation of large-scale galactic outflows, whereas chemical evolution models - which are often used for similar studies - are based on analytic prescriptions. Specifically, we do not assume instantaneous recycling of stellar mass loss nor instantaneous mixing of metals nor a specific parametrized gas accretion or star formation history. However, our simulations do not consider mixing of the gas between resolution elements.

In Section~\ref{sec:sim} we describe the suite of simulations used, as well as the way we compute the deuterium abundance and the fractional contribution of stellar mass loss to the gas, i.e.\ the `recycled gas fraction' or $f_\mathrm{recycled}$ (Section~\ref{sec:Dfrac}). The deuterium retention fraction and recycled gas fraction are related via $\mathrm{(D/D_{prim})}=1-f_\mathrm{recycled}$. In Section~\ref{sec:results} we present our results, including comparisons to existing observations. Section~\ref{sec:evol} describes the evolution of the deuterium fraction (and hence of the recycled gas fraction), while Section~\ref{sec:z3} focuses on high redshift and Section~\ref{sec:z0} on low redshift. We discuss our results and conclude in Section~\ref{sec:concl}.

\section{Method} \label{sec:sim}

The simulations used are part of the FIRE-1 sample. These were run with \textsc{gizmo}\footnote{http://www.tapir.caltech.edu/{\raise.17ex\hbox{$\scriptstyle\sim$}}phopkins/Site/GIZMO.html} \citep{Hopkins2015} in `P-SPH' mode, which adopts the Lagrangian `pressure-energy' formulation of the smoothed particle hydrodynamics (SPH) equations \citep{Hopkins2013PSPH}. The gravity solver is a heavily modified version of \textsc{gadget}-2 \citep{Springel2005}, with adaptive gravitational softening following \citet{Price2007}. Our implementation of P-SPH also includes substantial improvements in the artificial viscosity, entropy diffusion, adaptive timestepping, smoothing kernel, and gravitational softening algorithm. 

The FIRE project consists of a suite of cosmological `zoom-in' simulations of galaxies with a wide range of masses, simulated to $z=0$ (\citealt{Hopkins2014FIRE, Chan2015, Ma2016, Hafen2017}; Feldmann et al.\ in preparation), to $z=1.7$ \citep{Feldmann2016}, and to $z=2$ \citep{Faucher2015}. The simulation sample used is identical to the one used in \citet{Voortetal2016} and the simulation details are fully described in \citet{Hopkins2014FIRE} and references therein. The three Milky Way-mass galaxies that are the focus of Figure~\ref{fig:Devol} and~\ref{fig:Drad} are simulations `m12i', `m12v', and `m11.9a' (from highest to lowest stellar mass) from \citet{Hopkins2014FIRE} and \citet{Hafen2017}.
A $\Lambda$CDM cosmology is assumed with parameters consistent with the 9-yr Wilkinson Microwave Anisotropy Probe (WMAP) results \citep{Hinshaw2013}. The initial particle masses for baryons (dark matter) vary from $2.6\times10^2-4.5\times10^5$~M$_{\astrosun}$ ($1.3\times10^3-2.3\times10^6$~M$_{\astrosun}$) for the 16 simulations that were run to $z=0$ (see also \citealt{Voortetal2016} for further details). The 23 simulations that were run to $z\approx2$ are described in \citet{Faucher2015} and \citet{Feldmann2016} and their initial baryonic (dark matter) masses are $(3.3-5.9)\times10^4$~M$_{\astrosun}$ ($(1.7-2.9)\times10^5$~M$_{\astrosun}$). 

Star formation is restricted to molecular, self-gravitating gas above a hydrogen number density of $n_\mathrm{H}\approx5-50$~cm$^{-3}$, where the molecular fraction is calculated following \citet{Krumholz2011} and the self-gravitating criterion following \citet{Hopkins2013SelfGrav}. The majority of stars form at gas densities significantly higher than this imposed threshold. Stars are formed from gas satisfying these criteria at the rate $\dot\rho_\star=\rho_\mathrm{molecular}/t_\mathrm{ff}$, where $t_\mathrm{ff}$ is the free-fall time. When selected to undergo star formation, the entire gas particle is converted into a star particle.

We obtain stellar evolution results from STARBURST99 \citep{Leitherer1999} and assume an initial stellar mass function (IMF) from \citet{Kroupa2002}. Radiative cooling and heating are computed in the presence of the CMB radiation and the ultraviolet (UV)/X-ray background from \citet{Faucher2009}. Self-shielding is accounted for with a local Sobolev/Jeans length approximation. We impose a temperature floor of 10~K or the CMB temperature. 

The primordial abundances are $X_\mathrm{prim}=0.76$ and $Y_\mathrm{prim}=0.24$, where $X_\mathrm{prim}$ and $Y_\mathrm{prim}$ are the mass fractions of hydrogen and helium, respectively. The simulations include a metallicity floor at metal mass fraction $Z_\mathrm{prim}\approx10^{-4}$~Z$_{\astrosun}$ or $Z_\mathrm{prim}\approx10^{-3}$~Z$_{\astrosun}$, because yields are very uncertain at lower metallicities and we do not resolve the formation of individual first-generation stars. The abundances of 11 elements (H, He, C, N, O, Ne, Mg, Si, S, Ca and Fe) produced by massive and intermediate-mass stars are computed following \citet{Iwamoto1999}, \citet{Woosley1995}, and \citet{Izzard2004}. The amount of mass and metals ejected in a computational time-step depends on the age of the star particle and our simulations therefore self-consistently follow time-dependent chemical enrichment. Mass ejected through supernovae and stellar winds are modelled by transferring a fraction of the mass of a star particle to its neighbouring gas particles, $j$, within its SPH smoothing kernel as follows:
\begin{equation}
f_j = \dfrac{\frac{m_j}{\rho_j} W(r_j,h_\mathrm{sml})}{\Sigma_i \frac{m_i}{\rho_i} W(r_i,h_\mathrm{sml})},
\end{equation}
where $h_\mathrm{sml}$ is the smoothing length of the star particle (determined in the same manner as for gas particles), $r_i$ is the distance from the star particle to neighbour $i$, $W$ is the quintic SPH kernel, and the summation is over all SPH neighbours of the star particle, 62 on average. There is \emph{no} sub-resolution metal diffusion in these simulations.

The FIRE simulations include an explicit implementation of stellar feedback by supernovae, radiation pressure, stellar winds, and photo-ionization and photo-electric heating (see \citealt{Hopkins2014FIRE} and references therein for details). Feedback from active galactic nuclei (AGN) is not included. For star-forming galaxies, which constitute the majority of our simulated galaxies, AGN are thought to be unimportant. However, AGN-driven outflows are potentially important for the high-mass end of our simulated mass range. 

We measure a galaxy's stellar mass, $M_\mathrm{star}$, within 20~proper~kpc of its centre. The deuterium fraction of a galaxy's ISM is measured within 20~proper~kpc of its centre for gas with a temperature below $10^4$~K, which selects the warm ionized and cold neutral gas in the ISM. These choices have a mild effect on the normalization of some of our results, but not on the trends or on our conclusions.

\subsection{Deuterium fraction in hydrodynamical simulations} \label{sec:Dfrac}

Determining $\mathrm{(D/H)}/\mathrm{(D/H)}_\mathrm{prim}$ in our simulations is straightforward. The mass of a gas particle can only increase during the simulation by receiving mass lost from nearby stars (no particle splitting is implemented). Therefore, any mass above the initial particle mass, $m_\mathrm{initial}$, is deuterium-free. This is mixed with the initial particle mass, which has the primordial fraction of deuterium. Therefore, for each gas particle, we calculate
\begin{equation} \label{eqn:Dfrac}
\begin{split}
\frac{\mathrm{(D/H)}}{\mathrm{(D/H)}_\mathrm{prim}} & = \frac{\mathrm{D}}{\mathrm{D_{prim}}}\frac{\mathrm{H_{prim}}}{\mathrm{H}} = \frac{m_\mathrm{initial}}{m_\mathrm{initial}+m_\mathrm{recycled}}\frac{X_\mathrm{prim}}{X_\mathrm{gas}} \\
& = \frac{m_\mathrm{initial}}{m_\mathrm{gas}}\frac{X_\mathrm{prim}}{X_\mathrm{gas}}, 
\end{split}
\end{equation}
where $m_\mathrm{recycled}$ is the amount of mass received from evolving stars, i.e.\ the amount of gas that has been `recycled', $X_\mathrm{gas}$ is the mass fraction of hydrogen, and $m_\mathrm{gas}$ is the mass of the particle at the redshift of interest. We refer to this quantity as the deuterium retention fraction, because it is the fraction of deuterium, produced during Big Bang nucleosynthesis, that is not destroyed. The inverse of Equation~\ref{eqn:Dfrac} is the astration factor. The value of $\mathrm{(D/H)}_\mathrm{prim}$ is well-constrained, both directly from absorption-line observations of low-metallicity gas and indirectly from CMB measurements coupled with Big Bang nucleosynthesis reaction rates \citep[e.g.][]{Cooke2016, Coc2015, Cyburt2016}. Another way to constrain the primordial value is by comparing observations to cosmological simulations, as done in Section~\ref{sec:z3}.  

The deuterium retention fraction in Equation~\ref{eqn:Dfrac} is directly related to the fractional contribution of stellar mass loss to the gas, i.e.\ the recycled gas fraction,
\begin{equation} 
\begin{split}
f_\mathrm{recycled} & = \frac{m_\mathrm{recycled}}{m_\mathrm{gas}} = \frac{m_\mathrm{gas}-m_\mathrm{initial}}{m_\mathrm{gas}} \\
& = 1-\frac{\mathrm{D}}{\mathrm{D_{prim}}} = 1-\frac{\mathrm{(D/H)}}{\mathrm{(D/H)}_\mathrm{prim}}\frac{X_\mathrm{gas}}{X_\mathrm{prim}},
\end{split}
\end{equation}
which is used to study the importance of stellar mass loss in fuelling the ISM and star formation. Besides destroying all deuterium, a fraction of the hydrogen is fused into helium and metals before the gas is recycled into the ISM, $X_\mathrm{gas}=1-Y_\mathrm{gas}-Z_\mathrm{gas}$. This can be approximated well by $X_\mathrm{gas}\approx X_\mathrm{prim}-3Z_\mathrm{gas}$. The factor $X_\mathrm{gas}/X_\mathrm{prim}$ is close to unity for subsolar metallicities, but becomes more important at supersolar metallicities. Even though the differences do not change our conclusions, we will show and discuss both $\mathrm{(D/H)}/\mathrm{(D/H)}_\mathrm{prim}$ and $\mathrm{D/D_{prim}}$ or $f_\mathrm{recycled}$ when relevant.

\section{Results} \label{sec:results}

Observations of the deuterium fraction exist at both high and low redshift. We will first discuss the evolution of (D/H) and then discuss predictions and observational comparisons at $z=3$ and $z=0$ separately. Throughout the paper, we use oxygen abundance ratios of gas as compared to those of the Sun, i.e.\ $\mathrm{[O/H]}= \mathrm{log_{10}}(n_\mathrm{O}/n_\mathrm{H}) - \mathrm{log_{10}}(n_\mathrm{O}/n_\mathrm{H})_{\astrosun}$, where $n_\mathrm{O}$ is the oxygen number density, $n_\mathrm{H}$ the hydrogen number density, and $\mathrm{log_{10}}(n_\mathrm{O}/n_\mathrm{H})_{\astrosun}=-3.31$ is the solar oxygen abundance taken from \citet{Asplund2009}.

\subsection{Evolution of the deuterium fraction} \label{sec:evol}

\begin{figure*}
\center
\includegraphics[scale=.62]{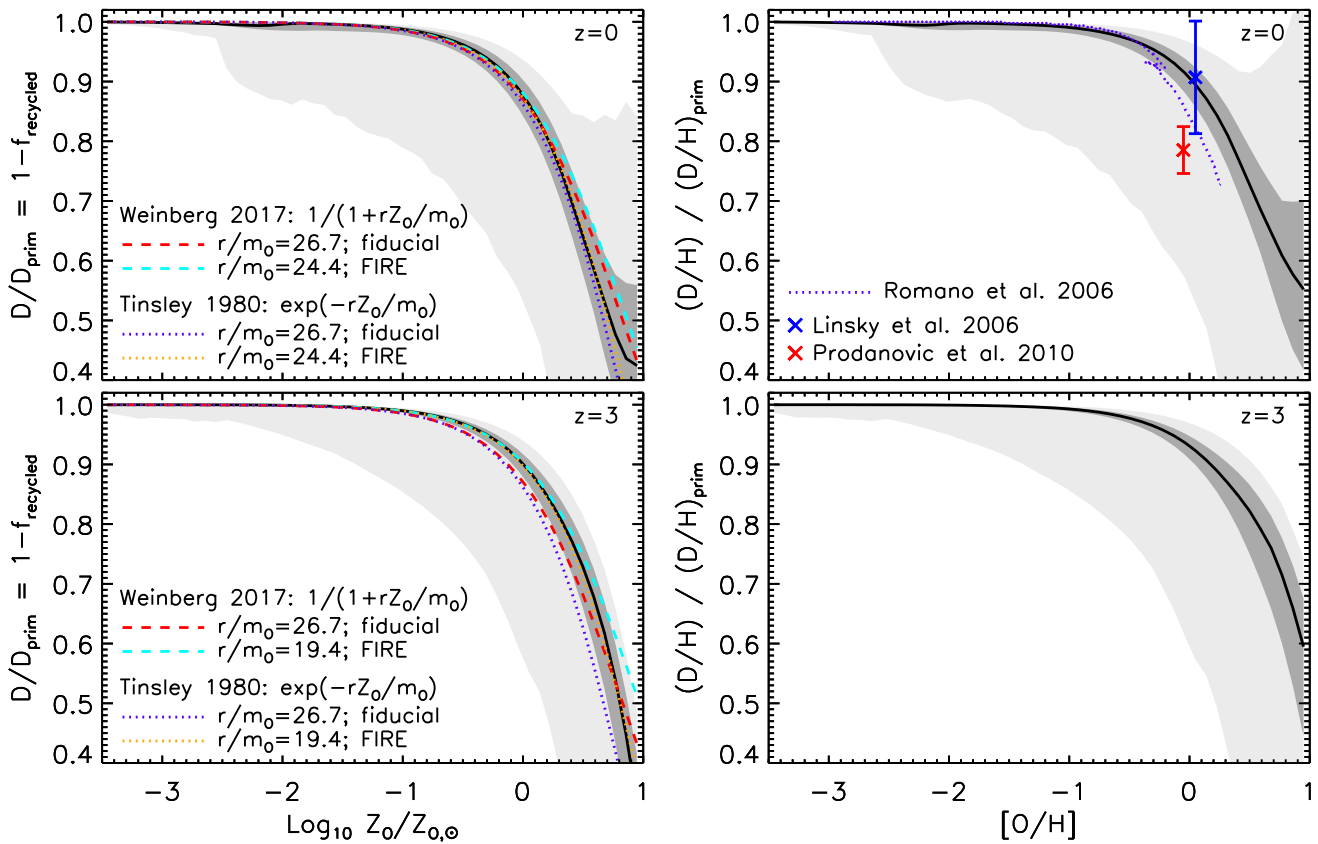}
\caption {\label{fig:DOz} The fraction of deuterium in all the gas particles in our simulations, normalized by the primordial deuterium fraction, as a function of oxygen metallicity at $z=0$ (top) and $z=3$ (bottom). We included all our simulations in the bottom panels and all the ones run down to $z=0$ in the top panels. The left panels only take into account the deuterium and oxygen abundance, whereas the right panels divide these by the hydrogen abundance, which decreases with increasing metallicity. The black curve shows the median deuterium retention fraction in our simulations and the grey shaded regions show the $1\sigma$ and $3\sigma$ scatter around the median. The dotted and dashed curves (left) show the relation derived from one-zone chemical evolution models \citep{Tinsley1980, Weinberg2017}. Our simulations match the chemical evolution models reasonably well and confirm that the deuterium fraction is close to primordial at $\mathrm{[O/H]}\lesssim-2$. The red, dashed and purple, dotted curve (left) assume fiducial values of $r=0.4$ and $m_\mathrm{O}=0.015$, while the cyan, dashed and orange, dotted curve use the average values extracted from our simulations at $z=0$ and $z=3$. The normalization of the deuterium-oxygen relation is slightly higher at $z=3$, when AGB stars have contributed less to the total stellar mass loss, reducing the ratio between the recycling fraction and the oxygen yield, which improves the match between the model and our simulations. A numerical chemical evolution model from \citet{Romano2006} is reproduced as a purple, dotted curve (right) and agrees qualitatively with our simulations, but shows tension at the $2\sigma$ level. This could be due to the different IMF and stellar evolution models used. We find that at solar oxygen metallicity, about 88 per cent of the primordial deuterium survives at $z=0$ and 90 per cent at $z=3$ (left panels). The values for $\mathrm{(D/H)}/\mathrm{(D/H)}_\mathrm{prim}$ (right panels) are slightly higher with $\mathrm{(D/H)}/\mathrm{(D/H)}_\mathrm{prim}=0.91$ ($z=0$) and $0.93$ ($z=3$) at $\mathrm{[O/H]}=0$ and agree with observational determinations of the deuterium fraction in the local Milky Way ISM \citep[crosses with error bars,][see text]{Linsky2006, Prodanovic2010}.} 
\end{figure*} 
While the total deuterium content of the Universe decreases with time, its total metallicity increases, leading to an inverse correlation between the deuterium and oxygen abundance \citep[e.g.][]{Steigman2003, Romano2006, Dvorkin2016}. Figure~\ref{fig:DOz} shows the median deuterium retention fraction (black curves) for all gas particles in our simulations as a function of oxygen metallicity at $z=0$ (top panels) and $z=3$ (bottom panels). The left panels only take into account deuterium and oxygen, whereas the right panels fold in the hydrogen abundance for both axes. $Z_\mathrm{O}$ ($Z_\mathrm{O, \astrosun}$) is the oxygen mass fraction of the gas (for solar abundances). The grey, shaded regions show the $1\sigma$ (dark) and $3\sigma$ (light) scatter around the median. Some of our simulations implemented a relatively high metallicity floor of $\mathrm{[O/H]}_\mathrm{initial}=-2.8$. Here, in order to not be affected by the imposed metallicity floor, we excluded gas with a metallicity within a factor of 2 from its initial oxygen abundance, but this choice does not affect our conclusions. At solar oxygen metallicity, about 90 per cent of the primordial deuterium is not destroyed at $z=3$ and 88 per cent at $z=0$. $\mathrm{(D/H)}/\mathrm{(D/H)}_\mathrm{prim}=0.91$ ($z=0$) and $0.93$ ($z=3$) at $\mathrm{[O/H]}=0$, slightly higher because of the small decrease of the hydrogen fraction. 

The $1\sigma$ scatter in this relation is very small, which shows that the destruction of deuterium and the enrichment with oxygen are tightly correlated. However, the scatter increases at $z=0$ at the highest metallicities ($\mathrm{[O/H]}\gtrsim0.5$). Additionally, we find large non-Gaussian tails at all metallicities, which means that even at low metallicity, a small fraction of gas particles have substantially reduced deuterium abundances. Our calculations likely underestimate the mixing of gas, because elements in our simulation are stuck to gas particles and do not diffuse to neighbouring gas particles. Adding turbulent diffusion to our simulations would only decrease the scatter in the correlation between deuterium and oxygen, because it smoothes out variations, and would thus strengthen our conclusions. The dependence of (D/H) on [O/H] is very steep at high metallicity, because [O/H] is a logarithmic quantity.

A small fraction of the gas (0.5 per cent) reaches extremely high metallicities ($\mathrm{[O/H]}>0.5$), which have not been observed. It is possible that such rare systems exist, outside the Milky Way, but are beyond current observational capabilities. Note, however, that the (average) metallicity in sightlines through our simulation is always $\mathrm{[O/H]}\leq0.5$ (see Figure~\ref{fig:HID}). Another possibility is that there is not enough mixing in our simulations, since the metals are stuck to particles and cannot diffuse, resulting in small metal-rich pockets. Additionally, the yields are very uncertain at such high metallicities. The real uncertainty is therefore larger than the scatter in this regime. Note that although $\mathrm{D/D_{prim}}\leq1$, $\mathrm{(D/H)}/\mathrm{(D/H)}_\mathrm{prim}$ can be larger than unity in rare cases at high metallicity, because $\mathrm{H/H_{prim}}$ can become very small due to hydrogen fusion. 

For comparison, the red, dashed curves (identical in top and bottom left panels) show the relation between the oxygen and deuterium abundances obtained from a one-zone chemical evolution model \citep{Weinberg2017}. This model assumes that chemical equilibrium is reached in the ISM due to the balance between gas inflow and outflow, enrichment though stellar mass loss, and gas consumption due to star formation. The only parameters in the relation are the recycling fraction, $r$, i.e.\ the fraction of mass returned to the ISM by a simple stellar population, and the oxygen yield, $m_\mathrm{O}$, i.e.\ the mass fraction of a simple stellar population released into the ISM in oxygen, 
\begin{equation} \label{eqn:W16}
\mathrm{\dfrac{D}{D_{prim}}}=\dfrac{1}{1+rZ_\mathrm{O}/m_\mathrm{O}},
\end{equation}
where $Z_\mathrm{O}$ is the oxygen mass fraction of the gas. We also compare our findings to analytic results derived from a closed box model, i.e.\ no gas inflow or outflow \citep{Tinsley1980}. The resulting relation between the deuterium retention fraction and oxygen abundance is\footnote{Note that in the literature, this equation is usually expressed with a yield defined with respect to the final stellar remnant mass (after mass loss) rather than the initial stellar mass as is the case in our definition of $m_\mathrm{O}$.} 
\begin{equation} \label{eqn:T80}
\mathrm{\dfrac{D}{D_{prim}}}=e^{\dfrac{-r Z_\mathrm{O}}{m_\mathrm{O}}}
\end{equation}
as shown by the purple, dotted curves (identical in top and bottom panels). Both chemical evolution models assume instantaneous stellar mass loss and enrichment, with no time dependence (whereas our simulations consistently follow time-dependent mass loss and enrichment as the stellar population ages). The ratio $r/m_\mathrm{O}=26.7$ (using $r=0.4$, and $m_\mathrm{O}=0.015$, the fiducial values from \citealt{Weinberg2017}) is thus the only free parameter. The models match the relative abundances at $z=0$ surprisingly well. 

As can be seen by comparing the two panels of Figure~\ref{fig:DOz}, there is relatively little evolution in the correlation between deuterium and oxygen. At fixed oxygen metallicity, the deuterium abundance is slightly higher at $z=3$ than at $z=0$. This is because most of the oxygen is produced in core-collapse supernovae, which also dominate the stellar mass loss at early times. At late times, AGB stars are responsible for most of the mass loss, adding deuterium-free material, but not substantially enriching the gas with oxygen. This can be tested by dividing the cumulative amount of mass loss added to the gas in the simulations by the total amount of gas-phase oxygen at different redshifts. As mentioned before, the fiducial ratio used is $r/m_\mathrm{O}=26.7$, which is close to, though slightly higher than, the value we find at $z=0$, $r/m_\mathrm{O}=24.4$ (directly computed from and averaged over all our simulations). At $z=3$, however, the average ratio in our simulations is substantially different, $r/m_\mathrm{O}=19.4$. We therefore added extra model curves (cyan, dashed using Equation~\ref{eqn:W16} and orange, dotted using Equation~\ref{eqn:T80}) to each panel of Figure~\ref{fig:DOz}, where we changed the value of $r/m_\mathrm{O}$ to match the average value in the simulations. 

The level of agreement between these simple models, especially the closed box model from \citet{Tinsley1980}, and our cosmological simulation results is remarkable given the very different approaches. This lends credence to both methods and shows that the most important factor in this correlation is the ratio $r/m_\mathrm{O}$, which can be calculated from stellar population synthesis models. The complex processes involved in the formation of galaxies, such as galaxy mergers, time-variable star formation and galactic outflows, as well as the lack of mixing in these simulations are thus likely unimportant where these relative abundances are concerned. The improvement from the small variation in $r/m_\mathrm{O}$ with redshift supports our claim that the evolution is due to the extra (almost oxygen-free) mass loss from AGB stars at late times.

Accurate observations of (D/H) at $\mathrm{[O/H]}>-1$ in combination with an accurate determination of $\mathrm{(D/H)}_\mathrm{prim}$, either from observations at low metallicity or derived from CMB measurements, would be able to determine $r/m_\mathrm{O}$. The recycling fraction is governed by intermediate-mass stars as well as massive stars, whereas the oxygen yield depends only on the latter. Therefore, the relation between $\mathrm{(D/H)/(D/H)}_\mathrm{prim}$ and [O/H] can potentially be used to constrain stellar evolution models and/or the variation of the IMF at the high-mass end. Although all our simulations were run with the same stellar evolution model and IMF, numerical chemical evolution models have already demonstrated that the deuterium fraction depends on these choices \citep[e.g.][]{Tosi1998, Prantzos2001, Romano2006}. 

In the top right panel, observational constraints independently derived by \citet{Linsky2006} (blue cross with error bars) and \citet{Prodanovic2010} (red cross with error bars) for the local ISM have been included, slightly offset from $\mathrm{[O/H]}=0$ for clarity and using $\mathrm{(D/H)}_\mathrm{prim}=2.547\times10^{-5}$, as recently obtained by \citet{Cooke2016}. Our solar deuterium value is in excellent agreement with that of \citet{Linsky2006}, but higher than that of \citet{Prodanovic2010}. The latter, however, is interpreted by the authors as a lower limit on the true value, in which case it is also in agreement with the result from our simulations (see Section~\ref{sec:z0} for further discussion on these results). 

Additionally, it is of interest to compare our results to those from numerical chemical evolution models. We therefore reproduced one of the models, based on a \citealt{Scalo1986} IMF and \citealt{Schaller1992} stellar lifetimes, from \citet{Romano2006}, shown in the top right panel as the purple, dotted curve. Although these results agree qualitatively, there is a clear $2\sigma$ discrepancy at high metallicity. This is potentially caused by the different IMF and stellar evolution models used, by the different star formation histories (and thus different importance of AGB stars), or by the inclusion of galactic outflows. Given the excellent match between our simulations and the simple closed box model, we believe the former explanation is the most plausible. 

\begin{figure}
\center
\includegraphics[scale=.52]{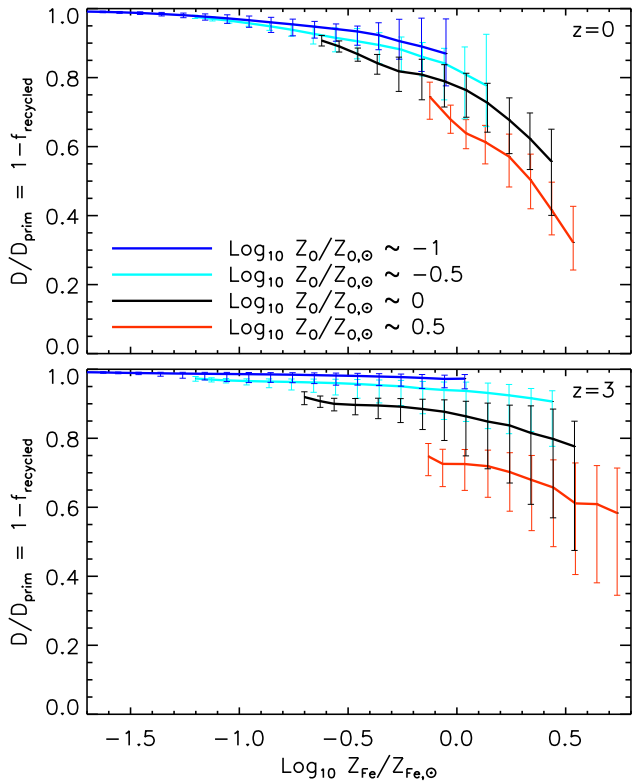}
\caption {\label{fig:DFe} The fraction of deuterium, normalized by the primordial deuterium fraction, as a function of iron metallicity at fixed oxygen metallicity at $z=0$ (top) and $z=3$ (bottom). The coloured curves show the median residual relation in four $Z_\mathrm{O}$ bins, 0.1 dex wide, centred on the value indicated by the legend and increasing from top to bottom. Error bars show the 16th and 84th percentiles of the distribution. Only bins containing at least 100 gas particles are shown. The deuterium fraction (recycled gas fraction) decreases (increases) with iron abundance, because intermediate-mass stars become relatively more important, ejecting iron-rich and deuterium-free material into the ISM. The residual correlation between deuterium and iron is stronger at lower redshift, due to the increased importance of older stars, which increases the scatter in the relation between deuterium and oxygen (Figure~\ref{fig:DOz}).}
\end{figure} 
Although the $1\sigma$ scatter in the relation between deuterium and oxygen abundances is small, it is nonzero and slightly larger at lower redshift. As mentioned before, almost all oxygen is produced by massive stars and released in core-collapse supernovae, whereas this is only the case for about half of the iron. The other half is synthesized in intermediate-mass stars and released in type Ia supernova explosions and winds from AGB stars \citep[e.g.][]{Wiersma2009b}. The iron abundance at fixed oxygen abundance therefore enables us to trace the relative importance of massive (younger) stars and intermediate-mass (older) stars and check whether variations in the contribution of stellar mass loss by AGB stars is responsible for the scatter seen in Figure~\ref{fig:DOz}. 

Figure~\ref{fig:DFe} shows the residual dependence of the normalized deuterium fraction on the iron metallicity at fixed oxygen metallicity at $z=0$ (top panel) and $z=3$ (bottom panel). $Z_\mathrm{Fe}/Z_\mathrm{Fe, \astrosun}$ is the iron mass fraction of the gas, normalized by the solar value. The coloured curves show the median relation between deuterium and iron in four bins, 0.1 dex wide, with (from top to bottom) $\mathrm{Log}_{10} Z_\mathrm{O}/Z_\mathrm{O, \astrosun}\approx-1$ (blue), $\approx-0.5$ (cyan), $\approx0$ (black), and $\approx0.5$ (red). Error bars show the 16th and 84th percentiles of the distribution and only bins containing at least 100~gas particles are included. 

A larger iron abundance at fixed $Z_\mathrm{O}/Z_\mathrm{O, \astrosun}$ means that older, intermediate-mass stars have been relatively more important for enriching the ISM. The deuterium fraction is thus expected to decrease with increasing iron abundance. Figure~\ref{fig:DFe} proves that this is indeed the case, although there is significant scatter in the residual relation between deuterium and iron. We therefore conclude that the (small) scatter in the relation between deuterium and oxygen is at least in part due to the varying importance of AGB stars and thus to the varying age of the stellar population responsible for enriching the gas. 

Comparing the $z=0$ and $z=3$ results, it is clear that the residual dependence of the deuterium fraction on the iron abundance is stronger at $z=0$ than at $z=3$. This is likely because the variation in stellar population ages, and thus in the importance of AGB stars, is smaller at higher redshift, when the Universe was much younger. This is also consistent with the fact that the $1\sigma$ scatter in the correlation between the deuterium fraction and oxygen abundance is smaller at higher redshift. Measuring $\mathrm{[Fe/H]}$ besides $\mathrm{[O/H]}$ and (D/H) will provide even better constraints on the stellar IMF and stellar evolution models.

\subsubsection{Milky Way-mass galaxies}

\begin{figure}
\center
\includegraphics[scale=.52]{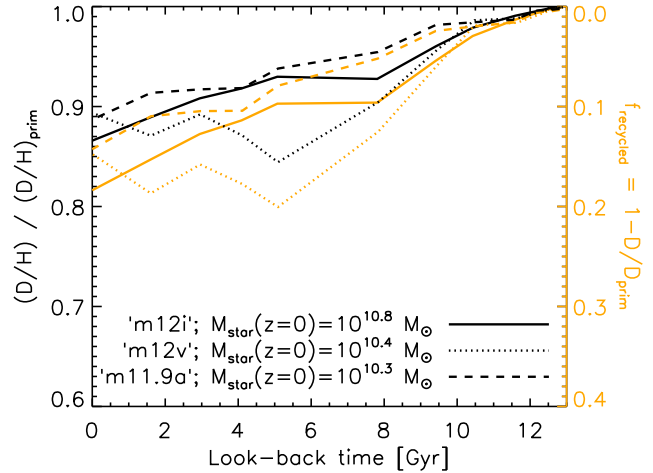}
\caption {\label{fig:Devol} Evolution of the mean fraction of deuterium in the ISM of three star-forming, Milky Way-mass galaxies, normalized by the primordial deuterium fraction. The black curves include the evolution of the hydrogen fraction (left axis), whereas the orange curves show the recycled gas fraction (right axis). The deuterium fraction in these galaxies decreases with time. For two of the galaxies (solid and dashed curves; `m12i' and `m11.9a'), no equilibrium value has been reached by $z=0$. This means that stellar mass loss becomes steadily more important for fuelling the ISM towards the present day. However, $\mathrm{(D/H)/(D/H)}_\mathrm{prim}$ levels out in the last $\approx5$~Gyr for one of the galaxies (dotted curve; `m12v'). This difference is likely related to their star formation history, because the majority of stellar mass loss occurs at young stellar ages. The former reach half of their present-day mass at $z\approx0.4$, whereas the latter already formed half of its stars by $z\approx1.1$.}
\end{figure}

To understand the chemical evolution of galaxies like the Milky Way, Figure~\ref{fig:Devol} shows the deuterium evolution for three of our simulated galaxies with stellar masses close to that of the Milky Way at $z=0$, as indicated in the legend. The mean deuterium retention fraction is calculated for the gas within 20~proper~kpc of the galaxies' centres, with a temperature below $10^4$~K, which selects the warm ionized and cold neutral gas in the ISM. The black curves include the evolution of the hydrogen fraction, as it would be measured in observations. The orange curves show the recycled gas fraction, which is lower, but show the same trends with look-back time. The final $z=0$ values vary between the three galaxies, because they have different stellar masses. The mass dependence will be discussed in Section~\ref{sec:z0}. Here, we are interested in the evolution of (D/H), that is, in the shape of the curves. Initially, the deuterium fraction is equal to its primordial value, after which it decreases. Two of the galaxies show an approximately linear decrease towards $z=0$ (solid and dashed curves), whereas for the galaxy with $M_\mathrm{star}=10^{10.4}$~M$_{\astrosun}$ (dotted curve; `m12v') the deuterium fraction levels off in the last $\approx5$~Gyr. The former have therefore not reached an equilibrium between the inflow of deuterium-rich gas from the IGM, the addition of deuterium-free gas through stellar mass loss, and the outflow of deuterium-poor gas. The latter galaxy has potentially reached chemical equilibrium in its ISM. 

Our three galaxies have different stellar masses and gas masses, on top of different star formation histories, and we lack the statistical power to control for this. Despite this limitation, we checked whether or not the low-redshift behaviour of the deuterium fraction is related to the star formation history. The galaxy which has reached deuterium equilibrium ($M_\mathrm{star}=10^{10.4}$~M$_{\astrosun}$; `m12v') has already formed half of its stars by $z\approx1.1$, whereas the other two reach half of their present-day stellar mass only at $z\approx0.4$ and have thus experienced much more low-redshift star formation and thus more low-redshift stellar mass loss. Therefore, the reason for the different low-redshift behaviour may indeed lie in the different star formation histories of our simulated galaxies.

We also checked for a dependence of the deuterium evolution on the mass loading factor, i.e.\ the gas outflow rate from a galaxy divided by the galaxy's SFR, as suggested by \citet{Weinberg2017}. The average mass loading factor at $z_1>z>z_2$ is calculated in the following way. We select all the gas particles within 20~proper~kpc of the galaxies' centres and a temperature below $10^4$~K. We then divide the total mass of those selected particles that have been turned into stars by $z=z_2$ by the total mass of the selected particles that are still gaseous, but located beyond 20~proper~kpc of the galaxies' centres at $z=z_2$. We take $z_1$ and $z_2$ to be approximately 1.5~Gyr apart, which is similar to the gas consumption time-scale. We find that the mass loading factor is relatively constant in the last 5~Gyr. From the most to least massive of our simulated Milky Way-mass galaxies, their average late-time mass loading factors over the last 5~Gyr are 0.2 (`m12i'), 0.4 (`m12v'), and 2.1 (`m11.9a'). The values for `m12i' and `m12v' are consistent with the upper limits from \citet{Muratov2015}, who argue that these low mass loading factors are not driven by galactic winds, but caused by random gas motions and/or close passages of satellite galaxies. We conclude that there is no clear correlation of the mass loading factor with the late-time deuterium evolution.

Knowing the evolution of (D/H) can thus potentially help us understand a galaxy's star formation history. This could be achieved for the Milky Way with an accurate determination of the deuterium fraction in giant planets in the Solar System, such as Jupiter, in combination with present-day measurements in the local ISM \citep{Lellouch2001}. The deuterium fraction in the giant planets provides a fossil record of the deuterium fraction in the local ISM during the time the Solar System was formed, about 4.5~Gyr ago. Using $\mathrm{(D/H)}_\mathrm{prim}=2.547\times10^{-5}$ from \citet{Cooke2016}, the measurement by \citet{Lellouch2001} implies $\mathrm{(D/H)/(D/H)}_\mathrm{prim}=0.82^{+0.12}_{-0.15}$ in Jupiter, which is consistent with all three of our simulated Milky Way-mass galaxies and is not precise enough to distinguish between a declining or constant deuterium fraction. Future observations with higher accuracy would be well-suited for this purpose.

\subsection{Deuterium fraction at high redshift} \label{sec:z3}

There has been a large observational effort to measure the deuterium fraction in metal-poor gas through absorption lines in spectra of background quasars. Lyman Limit Systems (LLSs; $10^{17.2}<N_\mathrm{H\,\textsc{i}}<10^{20.3}$~cm$^{-2}$, where $N_\mathrm{H\,\textsc{i}}$ is the $H\,\textsc{i}$ column density) and Damped Lyman-$\alpha$ Systems (DLAs; $N_\mathrm{H\,\textsc{i}}>10^{20.3}$~cm$^{-2}$) are optically thick to Lyman limit photons. To make a fair comparison between our simulations and these observations, we calculate column densities based only on the neutral gas. Because the gas comprising these strong absorbers is partially shielded from the ambient UV radiation, it is more neutral than if it were optically thin. This is taken into account in our simulations by using the fitting formula from \citet{Rahmati2013}, which has been shown to capture the effect of self-shielding well.

\citet{Cooke2014} argue that the most precise measurements can be made in absorbers with $N_\mathrm{H\,\textsc{i}}>10^{19}$~cm$^{-2}$. In order to compare to these systems, we also restrict ourselves to sightlines with column densities above this limit. Additionally, we discard the rare systems with $N_\mathrm{H\,\textsc{i}}>10^{21}$~cm$^{-2}$ in order to not be dominated by molecular gas. We note that neither this selection nor the self-shielding correction affects our results. We do not find a dependence of (D/H) on column density at fixed metallicity, so absorption line systems at any column density could be used. The vast majority of the selected high column density absorbers are located in the haloes around galaxies \citep{Voort2012}. We therefore use a simulated region of 300 by 300~proper~kpc centred on the main galaxy in each of our zoom-in simulations. We grid this volume into 1~by 1~proper~kpc pixels to calculate the column density of $\mathrm{H}\,\textsc{i}$, $\mathrm{D}\,\textsc{i}$, and $\mathrm{O}\,\textsc{i}$. We assume that the neutral fraction is the same for all three atoms, because their ionization potentials are very similar, as is also done in observations.

\begin{figure}
\center
\includegraphics[scale=.52]{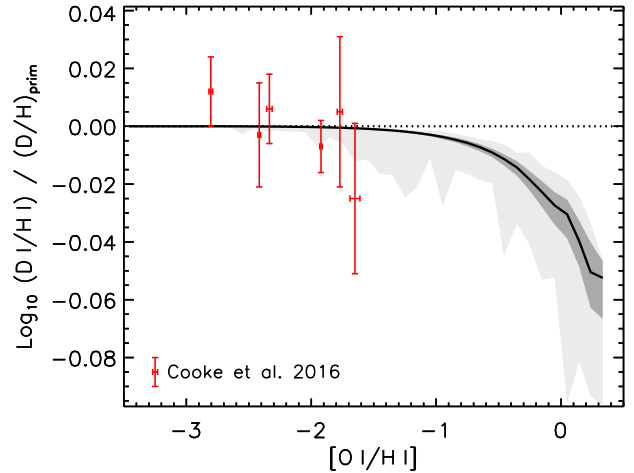}
\caption {\label{fig:HID} The fraction of deuterium in neutral gas in sightlines with column density $10^{19}<N_\mathrm{H\,\textsc{i}}<10^{21}$~cm$^{-2}$, normalized by the primordial deuterium fraction, as a function of oxygen metallicity at $z=3$. We include all our zoom-in simulations. The black curve shows the median deuterium retention fraction in our simulations and the grey shaded regions show the $1\sigma$ and $3\sigma$ scatter around the median. The red error bars show absorption-line observations, with associated $1\sigma$ measurement errors and assuming that $\mathrm{(D/H)}_\mathrm{prim}=2.547\times10^{-5}$ \citep{Cooke2016}. Comparing these measurements to our simulations, we determine a best-fit value for the primordial deuterium fraction of $\mathrm{(D/H)}_\mathrm{prim}=(2.549\pm0.033)\times10^{-5}$, consistent with the weighted mean of the measurements assuming no metallicity dependence. Because of the tight correlation between (D/H) and [O/H], more metal-rich absorbers could also be used for this purpose, by calibrating to the relation between deuterium and oxygen abundances found here, allowing for the expansion of the observational sample.}
\end{figure} 
The black curve in Figure~\ref{fig:HID} shows the median fraction of deuterium in neutral gas, divided by its primordial value, in the selected LLSs and DLAs at $z=3$ as a function of their metallicity. The different grey scales show the $1\sigma$ (dark) and $3\sigma$ (light) scatter around the median. Observations of $(\mathrm{D}\,\textsc{i}/\mathrm{H}\,\textsc{i})$ compiled by \citet{Cooke2016} and their associated $1\sigma$ errors are shown as red error bars, where we assumed that $\mathrm{(D/H)}_\mathrm{prim}=2.547\times10^{-5}$, the weighted mean of their measurements. 

Our simulations confirm that at $\mathrm{[O/H]}\lesssim-2$ the deuterium abundance is very close to the primordial value (within 0.1 per cent), as seen before in Figure~\ref{fig:DOz}. These low-metallicity systems are therefore appropriate to use to determine $\mathrm{(D/H)}_\mathrm{prim}$. At $\mathrm{[O/H]}=-1$ the median deuterium abundance is still only 1 per cent below primordial, similar to the $1\sigma$ error in the weighted mean of the observational values from \citet{Cooke2016}. The scatter in the relation is even smaller than in Figure~\ref{fig:DOz}, because we are including all (neutral) gas along a particular line-of-sight (rather than individual gas particles), decreasing the importance of small fluctuations. Observations of $N_\mathrm{H\,\textsc{i}}$, $N_\mathrm{D\,\textsc{i}}$, and $N_\mathrm{O\,\textsc{i}}$ are therefore well-suited to determine $\mathrm{(D/H)}_\mathrm{prim}$ and the relation between the deuterium and oxygen abundances. 

Instead of assuming no variation as a function of metallicity for the 6 observed systems shown in Figure~\ref{fig:HID}, we can test how well they match our simulations, which show a slight downward trend and minor additional scatter. We select those sightlines in our simulations that have the same metallicity as one of the observed absorbers, within $1\sigma$ errors. We then use least square fitting and calculate $\chi^2$ between our simulated sightlines and the observations as a function of $\mathrm{(D/H)}_\mathrm{prim}$, which sets the relative normalization. The minimum $\chi^2$ is reached for $\mathrm{(D/H)}_\mathrm{prim}=(2.549\pm0.033)\times10^{-5}$, where the errors are $1\sigma$ and calculated from the difference in $\chi^2$. This is consistent with theoretical models of Big Bang nucleosynthesis, based on cosmological parameters \citep{Coc2015, Cyburt2016}. Our best estimate is very similar to, though slightly higher than, the weighted mean calculated by \citet{Cooke2016}, who assumed no metallicity dependence. For this low-metallicity sample, we do not gain much accuracy from comparing the data to our simulations.

However, given that the scatter in the simulations is much lower than the observational measurement error at all metallicities, more metal-rich absorption-line systems can be used to determine the primordial deuterium fraction. This would allow for the expansion of the observational sample, which would improve the accuracy of $\mathrm{(D/H)}_\mathrm{prim}$. Even absorbers with $\mathrm{[O/H]}\gtrsim-1$ can be used when taking into account the relation between the deuterium and oxygen abundance based on hydrodynamical simulations or on Equation~\ref{eqn:T80} combined with a prescription for the change of $X_\mathrm{gas}$ with metallicity. For the latter, one should use a slowly evolving ratio of recycling fraction to oxygen yield, $r/m_\mathrm{O}$, increasing with time as the contribution of mass lost by intermediate-mass stars increases. Vice versa, observations of metal-rich absorbers can set constraints on the ratio of the recycling fraction and the oxygen yield, assuming that the primordial abundance of deuterium is known from either CMB measurements or from absorption-line observations at $\mathrm{[O/H]}\lesssim-2$. $r/m_\mathrm{O}$ depends on the relative number of intermediate- and high-mass stars and on their stellar yields and can thus potentially help constrain the high-mass end of the stellar IMF and/or stellar evolution models. 

It is important to note that the depletion of deuterium onto dust and preferential incorporation into molecules could cause large scatter in (D/H) between quasar sightlines at fixed metallicity, which are not due to variations in the recycled gas fraction. This is probably seen in the local ISM at solar metallicity \citep[e.g.][]{Wood2004, Linsky2006, Prodanovic2010} and briefly discussed in Section~\ref{sec:z0}. Unfortunately we cannot address this issue with our current simulations. A relatively large sample of (D/H) measurements in absorption-line systems could quantify the scatter in (D/H) between sightlines at fixed metallicity. This will tell us whether the depletion of deuterium onto dust is important in the intergalactic medium at $\mathrm{[O/H]}>-2$, because our simulations have shown that the scatter due to variations in stellar mass loss at fixed metallicity is negligibly small. If dust depletion turns out to be dominant, these systems cannot be used for determining $\mathrm{(D/H)}_\mathrm{prim}$ or $r/m_\mathrm{O}$. However, there is some evidence that LLSs tend to reside in dust-poor environments \citep{Fumagalli2016}. Additionally, dust depletion seems to be less important in lower column density absorbers \citep[e.g][]{Linsky2006, Prodanovic2010}. It is therefore possible that these systems are well-suited for determining $\mathrm{(D/H)}_\mathrm{prim}$ or $r/m_\mathrm{O}$ even at $\mathrm{[O/H]}>-2$.

\subsection{Deuterium fraction at low redshift} \label{sec:z0}

\begin{figure}
\center
\includegraphics[scale=.52]{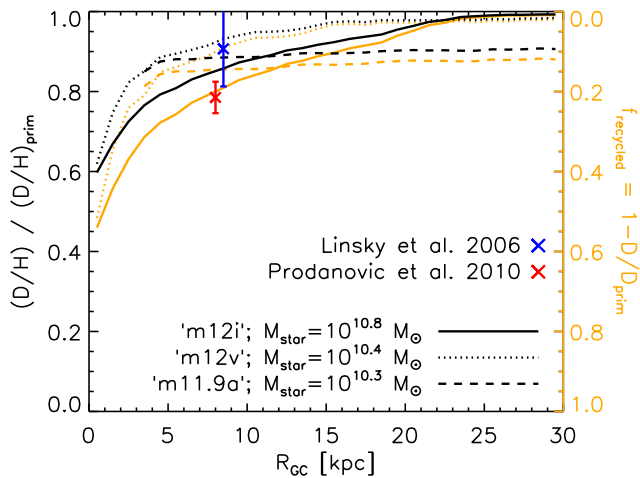}
\caption {\label{fig:Drad} The mean fraction of deuterium in the ISM of three star-forming Milky Way-mass galaxies, normalized by the primordial deuterium fraction, as a function of galactocentric radius at $z=0$ (black curves; left axis). The radial dependence of the recycled gas fraction is similar to that of the deuterium fraction and shown as orange curves (right axis). One galaxy (dashed curve; `m11.9a') has a central hole in its ISM at $R_\mathrm{GC}<4$~kpc, created by a strong outflow. This galaxy also has the lowest deuterium abundance at large radii, because it ejected a large amount of gas from its centre into its surroundings. For all three galaxies, the deuterium fraction increases with galactocentric radius. Our simulations are consistent with observational determinations of the deuterium fraction in the local Milky Way ISM \citep[crosses with error bars,][]{Linsky2006, Prodanovic2010}. In the galaxy centres, where most star formation occurs, about half of the gas originates from stellar mass loss.}
\end{figure}

To compare with observations of (D/H) at $z=0$, we focus on the deuterium fraction in the ISM of our simulated Milky Way-like galaxies. The black curves in Figure~\ref{fig:Drad} show how the ratio of the present-day abundance of deuterium to the primordial abundance varies with 3D distance from the galactic centre, $R_\mathrm{GC}$, for the same three galaxies as shown in Figure~\ref{fig:Devol}. For completeness, we show the recycled gas fraction as orange curves. $\mathrm{(D/D_{prim})}$ is similar to, though slightly lower than,  $\mathrm{(D/H)/(D/H)}_\mathrm{prim}$, because of the decrease of the hydrogen fraction. One galaxy (dashed curve; `m11.9a') has a central hole in its ISM, created by galactic winds, consistent with its relatively large average mass loading factor (see Section~\ref{sec:evol}). It therefore has no deuterium measurement at $R_\mathrm{GC}<4$~kpc. This galaxy has the lowest deuterium abundance at large radii, because the gas that was originally in its centre has been moved to larger radii. The deuterium retention fraction is low in the centres of the other two galaxies, where the density of stars is high and most of the star formation takes place. The deuterium fraction for all three galaxies increases with galactocentric radius, as previously shown by chemical evolution models \citep[e.g.][]{Prantzos1996, Chiappini2002, Romano2006}. The importance of stellar mass loss therefore increases towards the galaxy centre and recycled gas accounts for about half of the gas in the central kpc. The steepness of the deuterium abundance gradient could also reveal information on the assembly history of a galaxy \citep[e.g.][]{Prantzos1996}. In our sample, the galaxy with the flattest deuterium profile has the highest outflow rate. It may therefore depend more strongly on (bursty) galactic outflows than on (smooth) gas accretion. 

Large scatter exists between measurements of the deuterium abundance in the local ISM via absorption-line observations. This scatter could be explained by localized infall of pristine gas, with very little mixing. In this case, the average astration factor is relatively high (and mass lost from stars dominates the ISM). However, if this is the case, the oxygen abundance is also expected to decrease locally as it becomes diluted with the metal-free, infalling gas, resulting in large scatter. The fact that oxygen shows much smaller abundance variation than deuterium argues against such localized infall \citep{Oliveira2005}. Another, more likely, explanation for the large (D/H) sightline variations is that some of the deuterium is depleted onto dust. The probability of deuterium depletion onto dust grains and incorporation into molecules is high, since the zero-point energies of deuterium-metal bonds are lower than those of the corresponding hydrogen-metal bonds \citep{Jura1982, Tielens1983}. When the ISM is heated, dust grains and molecules can be destroyed, returning deuterium to the atomic gas phase. Metals, such as iron, silicon, and titanium, are also depleted onto dust grains and the correlation of their abundances with deuterium supports this theory \citep{Prochaska2005, Linsky2006, Lallement2008}. Based on the assumption that the observational scatter is caused by deuterium depletion onto dust, relatively high deuterium abundances, and low astration factors, are derived for the local Milky Way ISM by \citet{Linsky2006} and \citet{Prodanovic2010}. 

The deuterium retention percentages in the solar neighbourhood, here defined as $7<R_{GC}<9$~kpc, lie between 85 and 92 per cent for our three simulations of star-forming galaxies with masses similar to that of the Milky Way\footnote{Although our calculations are done using 3D distance, the results are unchanged when we restrict ourselves to the ISM, because most of the gas mass lies within the star-forming disc.}. This is consistent with Figure~\ref{fig:DOz}, where we found that 91 per cent of its primordial value is recovered for $\mathrm{[O/H]}=0$ at $z=0$. Using the value from \citet{Cooke2016} for the primordial deuterium fraction as in Section~\ref{sec:z3}, the deuterium abundance derived by \citet{Linsky2006} implies that the local ISM still contains $91^{+9}_{-10}$ per cent of the primordial deuterium abundance. This is consistent with our simulations within $1\sigma$. \citet{Prodanovic2010} use the same data compilation, but a different method, to derive a deuterium retention percentage of $79\pm4$ per cent (again assuming $\mathrm{(D/H)}_\mathrm{prim}=2.547\times10^{-5}$). This is consistent (within $2\sigma$) with our most massive Milky Way-like galaxy, with $M_\mathrm{star}=10^{10.8}$~M$_{\astrosun}$. However, \citet{Prodanovic2010} stress that their measurement can also be interpreted as a lower limit in the event that all available sightlines are affected by dust depletion. In this case, our other galaxies are also consistent with their model. Our simulations exhibit low astration factors and therefore agree with the explanation that the large scatter in local ISM observations is due to dust depletion rather than due to poor mixing of freshly accreted gas.  

\begin{figure}
\center
\includegraphics[scale=.52]{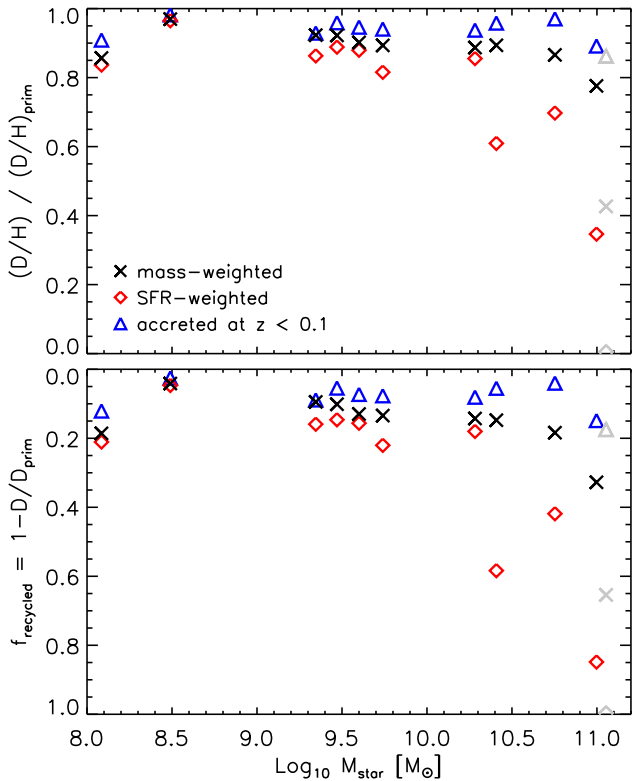}
\caption {\label{fig:DMstar} The abundance of deuterium normalized by the primordial deuterium abundance (top panel) and the recycled gas fraction (bottom panel) in each galaxy's ISM, as a function of stellar mass at $z=0$. The black crosses show the mass-weighted mean deuterium retention fraction and the red diamonds show the SFR-weighted mean values. The blue triangles show the mass-weighted $\mathrm{(D/H)/(D/H)}_\mathrm{prim}$ (or $f_\mathrm{recycled}$) for the gas that accreted, i.e.\ reached $R_\mathrm{GC}<20$~kpc and $T<10^4$~K, after $z=0.1$. The most massive galaxy (grey symbols) has limited ISM resolution and therefore large uncertainty in its deuterium abundance. In general, the deuterium fraction decreases with increasing stellar mass. This means that stellar mass loss is more important for feeding the ISM of high-mass galaxies. Stellar mass loss is even more critical for fuelling star formation, which preferentially occurs at the highest ISM densities. Accreting gas has a deuterium fraction between that of the ISM and the primordial value, which likely means that it is a combination of pristine infalling gas and gas ejected or stripped from satellites and/or gas reaccreting as part of a galactic fountain.}
\end{figure}

No known galaxy besides the Milky Way has a measurement of the deuterium fraction in their ISM. Such observations would be interesting, because our simulations predict a strong dependence on stellar mass. Figure~\ref{fig:DMstar} shows the mean deuterium retention fraction (top panel) and recycled gas fraction (bottom panel) within 20~kpc of the centre of the galaxy for gas with a temperature below $10^4$~K as a function of stellar mass at $z=0$. Due to the depletion of hydrogen at higher metallicity, the differences between the deuterium retention fraction and recycled gas fraction increase with stellar mass, but the trends with mass remain the same. The black crosses show the mass-weighted mean, while the red diamonds show the (instantaneous) SFR-weighted mean. The latter is therefore a better indicator of how important stellar mass loss is for the fuelling of star formation, whereas the former is the value that would be measured, for example, in sightlines through the ISM. The galaxy shown in grey with the highest stellar mass has large uncertainties for its deuterium abundance, because it contains only 12 star-forming gas particles (but more than 1000 gas particles in total). Two other massive galaxies are not included, because they contained no star-forming gas and very little non-star-forming gas. Galaxies with $M_\mathrm{star}<10^8$~M$_{\astrosun}$ are also excluded, because they contained no or very little star-forming gas. The other galaxies (shown in black, red, and blue) have 100 star-forming gas particles or more. 

There is a definite trend with stellar mass, where more massive galaxies generally have lower deuterium fractions and thus a larger contribution of stellar mass loss. This is a clear prediction from our simulations. The exception is the lowest mass galaxy which has a low (D/H), as well as a high [O/H] (as expected from Figure~\ref{fig:DOz}). This reflects the variation in star formation, inflow, and outflow history. Our sample of galaxies is limited, but quantifying the scatter in the deuterium fraction between galaxies of similar stellar mass would help us to understand how this correlates with their formation history. 

Due to their tight relation, the oxygen abundance could provide very similar information as the deuterium abundance. We expect this to work well at subsolar metallicities. At solar and supersolar metallicities, the dependence of (D/H) on [O/H] is very steep, which means that to determine the recycled gas fraction, very precise measurements of [O/H] are needed. Additionally, the scatter in the relation increases towards high metallicity at $z=0$, due to varying contributions of AGB stars, impeding the determination of the recycled gas fraction through [O/H]. 

The deuterium fraction is always lower when weighted by star formation rate (red diamonds) rather than by mass (black crosses), because this dense, star-forming gas is close to newly formed stars with high mass loss rates. This means that mass loss is more important for fuelling star formation than it is for replenishing the more diffuse ISM. A measurement of the deuterium fraction (such as those in the local ISM) generally provides information on the latter, but not the former. 

In our simulations, the star formation in low-mass galaxies is fuelled predominantly by gas accretion, although there is a small contribution by stellar mass loss. For two of the galaxies with $M_\mathrm{star}\approx10^{10.5}$~M$_{\astrosun}$, mass loss fuels about half of the current star formation, while one other is still dominated by gas accretion. For the massive galaxy (or galaxies, if we include the one with large uncertainties) with $M_\mathrm{star}\approx10^{11}$~M$_{\astrosun}$, mass loss fuels the majority of star formation. As mentioned in Section~\ref{sec:intro}, stellar mass loss has been suggested as fuel for the observed star formation in local, massive, early-type galaxies and in central cluster galaxies. Our simulations are consistent with this interpretation. 

Notably, there are also detections of neutral deuterium in quasar sightlines through clouds outside the galactic disc of the Milky Way, in the lower galactic halo \citep{Sembach2004, Savage2007}. These clouds could provide the fuel to sustain the Milky Way's steady star formation. However, the errors associated with these measurements are large and therefore not very constraining. More precise determinations of (D/H) could help constrain the nature of this gas. If the deuterium fraction of the gas around the Milky Way is high, it is likely pristine gas falling in from the IGM. If the deuterium fraction is similar to that of the Milky Way, it is likely part of a galactic fountain \citep{Shapiro1976}. In intermediate cases, the gas could be a mix of both high and low (D/H) gas or the gas could be ejected or stripped from lower mass satellites \citep[e.g.][]{Angles2017}. These different theories can potentially be tested via estimates of the scatter in (D/H) in the halo, which would be large if the fountain gas is not fully mixed with the gas falling in from the IGM, but small if the gas was ejected or stripped from satellites.

In order to compare the deuterium fraction in the galaxy to the deuterium fraction of accreting gas, the blue triangles in Figure~\ref{fig:DMstar} show $\mathrm{(D/H)/(D/H)}_\mathrm{prim}$ at $z=0.1$ of the gas that is accreted at $0<z<0.1$, which means it is located at $R_\mathrm{GC}<20$~kpc and has $T<10^4$~K at $z=0$, but is located at larger radii and/or has higher temperatures at $z=0.1$. Our radial boundary is somewhat arbitrary, but our results are not very sensitive to this (except for the normalization, because the deuterium fraction is generally higher at larger radius). The accreting gas has a higher deuterium retention fraction than the ISM gas. This is consistent with a substantial part of the material accreting for the first time onto a galaxy. The fact that $\mathrm{(D/H)/(D/H)}_\mathrm{prim}<1$, however, shows that another part of the accreting gas has previously been ejected or stripped from a galaxy. Such a combination of different accretion channels has already been found in these and other simulations \citep[e.g.][]{Oppenheimer2008, Christensen2016, Angles2017, Voort2017}. The difference in (D/H) between accreting gas and the ISM is especially large for the more massive galaxies in our sample. As mentioned above, this balance between accretion from the IGM and reaccretion could potentially be shown observationally by measuring (D/H) in clouds that are accreting onto the Milky Way \citep[e.g.][]{Sembach2004}.

\section{Discussion and conclusions} \label{sec:concl}

We have quantified the evolution of the deuterium fraction and its dependence on the oxygen abundance, galactocentric radius, and stellar mass in cosmological zoom-in simulations with strong stellar feedback. Because deuterium is only synthesized in the early Universe, it provides an interesting constraint on cosmology and on galaxy evolution. The normalized deuterium fraction is a measure of the recycled gas fraction, i.e.\ the fractional contribution of stellar mass loss to the gas, because deuterium is completely destroyed in stars and therefore $f_\mathrm{recycled}=1-\mathrm{(D/D_{prim})}$. Observations, however, measure $\mathrm{(D/H)/(D/H)}_\mathrm{prim}$, which is higher than $\mathrm{(D/D_{prim})}$, especially at high metallicity (because $X_\mathrm{gas}/X_\mathrm{prim}\approx1-3Z_\mathrm{gas}$). Our simulations self-consistently follow gas flows into and out of galaxies and the (metal-rich and deuterium-free) mass loss by supernovae and AGB stars. We have compared our predictions to available observations at low and high redshift and found them to be consistent. 
Our main conclusions can be summarized as follows:
\begin{itemize}
\item The deuterium fraction exhibits a tight correlation with the oxygen abundance, evolving slowly with redshift (Figure~\ref{fig:DOz}). This is captured well by simple chemical evolution models \citep{Tinsley1980,Weinberg2017}, which depend only on the ratio of the recycling fraction and the oxygen yield. We find a small increase in $r/m_\mathrm{O}$ with time, because of the increased importance of AGB stars. The variation in the importance of AGB stars at fixed oxygen abundance can be traced by the iron abundance and is responsible for some of the (small) scatter in the deuterium fraction (Figure~\ref{fig:DFe}).
\item The three Milky Way-mass galaxies in our sample exhibit different evolution at low redshift (Figure~\ref{fig:Devol}). The galaxies that form many of their stars at late times have continually decreasing deuterium fractions in their ISM. The galaxy which forms most of its stars before $z=1$ shows a constant deuterium fraction in the last $\approx5$~Gyr, indicating that it may have reached chemical equilibrium. The evolution of the deuterium fraction may therefore be directly connected to the galaxy's star formation history.
\item The deuterium fraction is very close to primordial at $\mathrm{[O/H]}\lesssim-2$ (within 0.1 per cent). These are the metallicities of LLSs and DLAs typically used to measure the primordial deuterium abundance. Because of the tight correlation with metallicity, deuterium measurements in more metal-rich systems can also be used to constrain the primordial deuterium fraction (Figure~\ref{fig:HID}) if dust depletion is unimportant.
\item We compared our simulations to the observational sample of \citet{Cooke2016} to determine a primordial deuterium fraction of $\mathrm{(D/H)}_\mathrm{prim}=(2.549\pm0.033)\times10^{-5}$, very close to, though slightly higher than, their original estimate, which assumed no dependence of the measured (D/H) on metallicity. Our result is also in agreement with cosmological parameters and Big Bang nucleosynthesis \citep{Coc2015, Cyburt2016}.
\item The deuterium fraction increases with galactocentric radius. Our simulations are consistent with the available estimates from the local Milky Way ISM where the observed scatter between sightlines is assumed to be caused by depletion onto dust (Figure~\ref{fig:Drad}).
\item The deuterium fraction decreases with increasing stellar mass, which means that the importance of stellar mass loss in our simulations increases with stellar mass (Figure~\ref{fig:DMstar}). Mass loss is more important for fuelling star formation than for replenishing the general ISM (which has, on average, a lower gas density). Accreting gas has a higher deuterium fraction than the ISM of galaxies, but lower than primordial. This is consistent with previous findings that some gas accretes directly from the IGM, but some has been previously been ejected or stripped from a galaxy \citep[e.g.][]{Angles2017}.
\end{itemize}

Due to the tight correlation of (D/H) and [O/H], shown in Figure~\ref{fig:DOz}, measurements of the oxygen abundance could provide the same information as the deuterium abundance if this relation is accurately calibrated by observations at $\mathrm{[O/H]}\gtrsim-1$. The relation evolves slowly, because of the increased importance of stellar mass loss from AGB stars, which should be taken into account (see Section~\ref{sec:evol}). If the importance of mass loss for fuelling the ISM increases for massive galaxies, as in our simulations, (D/H) will decrease with mass and [O/H] (and other metal abundances) will increase. There is observational evidence from gas-phase and stellar metallicities that this is indeed the case, although [O/H] may saturate at the highest stellar masses \citep[e.g.][]{Tremonti2004, Gallazzi2005, Mannucci2010, Peng2015}. Using the median relation derived from our simulations (Figure~\ref{fig:DOz}) one can immediately estimate the contribution of stellar mass loss given the gas-phase oxygen metallicity. However, this only works in the situation where galactic winds remove mass loss from supernovae and AGB stars (approximately) equally, but may not if ejecta from young stars are removed from a galaxy (through supernovae or AGN, quenching star formation) after which its ISM is replenished by mass loss from old stars alone. This is likely the reason that the scatter in (D/H) increases at supersolar metallicity in our simulations. In observations, however, the scatter at high metallicity is probably dominated by the depletion of deuterium onto dust, an effect which is not included in our calculations. 

Our cosmological, hydrodynamical simulations follow time-dependent chemical enrichment and the assembly of galaxies self-consistently, whereas simple, analytical chemical evolution models assume instantaneous recycling and specify a specific star formation history \citep{Tinsley1980,Weinberg2017}. We nevertheless compare our results to these models and find a remarkable agreement (especially with \citealt{Tinsley1980}) when considering the relation between the deuterium and oxygen abundance, despite the very different methods used. Although other galaxy properties are not necessarily well reproduced in these simplified models \citep[e.g.][]{Binney1998}, these do not play a dominant role when relative abundances are concerned. The bursty star formation and strong galactic outflows present in our simulations thus have no major impact on the correlation between (D/H) and [O/H]. Although \citet{Weinberg2017} find that a relatively high mass loading factor is necessary to match the $z=0$ deuterium abundance in the local ISM, two of our simulated galaxies show low mass loading factors at late times (but high mass loading factors at early times, see \citealt{Muratov2015}).  

\citet{Leitner2011} use zoom-in simulations with and without stellar mass loss to show that mass loss dominates the fuelling of star formation at late times for galaxies in haloes of similar mass to that of the Milky Way. However, their simulation did not include outflows from either star formation or AGN, which results in the galaxies being too massive at $z=0$. This means that too much baryonic mass is locked up in stars and therefore less gas is available for accretion at late times. Furthermore, the majority of the mass lost by stars is retained by the galaxy and not ejected by a galactic wind. Our simulations show that for Milky-Way mass galaxies with strong stellar feedback, cosmological inflow either dominates over or rivals stellar mass loss for the fuelling of star formation. However, recycled gas dominates the SFR at $M_\mathrm{star}\approx10^{11}$~M$_{\astrosun}$ in our simulations. 

\citet{Segers2016} use large-volume simulations with stellar and AGN feedback that match the stellar--to--halo mass relation and find that the contribution of mass loss to the fuelling of star formation is largest for galaxies with $M_\mathrm{star}\approx10^{10.5}$~M$_{\astrosun}$. The deuterium fraction in the ISM is therefore the lowest around this mass and increases for more massive galaxies. Our simulations (without AGN feedback) find a different behaviour at the high-mass end, where mass loss becomes increasingly important and the deuterium fraction decreases. \citet{Segers2016} also show results from simulations without AGN feedback, which do not match the stellar--to--halo mass relation, but agree qualitatively with our simulations. In the absence of AGN feedback, there are no strong galactic outflows, which means that most of the stellar mass loss is retained in the ISM. Different implementations of AGN feedback (ejective or preventive) could also result in different deuterium abundances. Future observations of (D/H) in massive galaxies therefore have the potential to discriminate between these different models. 

Numerical chemical evolution models generally find somewhat lower deuterium fractions (higher astration factors) than we do in our cosmological simulations \citep[e.g.][]{Dearborn1996, Tosi1998, Chiappini2002, Romano2006}. This mild difference could be due to the different assumptions made for the stellar IMF, metal yields, and stellar mass loss. A detailed quantitative comparison between the two different numerical approaches requires using the same IMF and stellar evolution models, which is left for future work. The star formation history of a galaxy determines the relative importance of AGB stars with respect to core-collapse supernovae, which may explain any remaining discrepancy. This could potentially be tested by comparing the oxygen and iron abundances (see Figure~\ref{fig:DFe}). Another possibility is that the inclusion of galactic outflows in our simulations reduces the importance of stellar mass loss, because a substantial fraction of the recycled gas is ejected from the ISM. 

The large observed scatter in the deuterium fraction between different sightlines through the local ISM could be interpreted as evidence for an inhomogeneous ISM due to localized gas accretion, giving rise to a low average deuterium fraction \citep[e.g.][]{Hebrard2003}. Our work instead favours the interpretation that the ISM is well-mixed and the observational scatter is caused by the depletion of deuterium onto dust, which leads to a local deuterium fraction not much lower than the primordial value \citep{Linsky2006, Prodanovic2010}. It is also possible that both dust depletion and localized infall play a roll and the true value is intermediate between those cases \citep{Steigmanetal2007}. More precise observations of the deuterium and metal abundances would help clarify which of these interpretations is correct.

In summary, we have quantified the deuterium fraction in a suite of zoom-in simulations and found it to be tightly correlated with the oxygen metallicity and consistent with current observational constraints. We conclude that the primordial deuterium fraction (and thus early cosmological expansion and Big Bang nucleosynthesis) can also be constrained by using observations at medium to high metallicity in combination with our simulations. Or, vice versa, if the primordial deuterium fraction is known, these measurements can inform us about the ratio of the recycling fraction to the oxygen yield and thus about the high-mass end of the stellar IMF and stellar evolution models. Our simulations predict that the deuterium fraction is lower at smaller galactocentric radii and for higher mass galaxies. This means that stellar mass loss could provide most of the fuel for star formation in massive early-type galaxies and in the centres of less massive, star-forming galaxies. Grid-based calculations or SPH simulations with explicit diffusion would be useful to determine whether or not small-scale mixing modifies the deuterium and oxygen abundances. Accurate observations of the deuterium fraction provide us with the possibility to understand the fuelling of star formation through stellar mass loss in galaxies in general and the Milky Way in particular.

\section*{Acknowledgements}

We would like to thank the Simons Foundation and the organizers and participants of the Simons Symposium `Galactic Superwinds: Beyond Phenomenology', in particular David Weinberg, for interesting discussions and inspiration for this work. We also thank Thomas Guillet and Joop Schaye for helpful discussions and Tim Davis for useful comments on an earlier version of the manuscript. We would like to thank the referees for valuable comments that helped clarify our results and put them into context.
Support for FvdV was provided by the Klaus Tschira Foundation.
EQ was supported by NASA ATP grant 12-APT12-0183, a Simons Investigator award from the Simons Foundation, and the David and Lucile Packard Foundation.
CAFG was supported by NSF through grants AST-1412836 and AST- 1517491 and by NASA through grant NNX15AB22G.
DK was supported by the NSF through grant AST-1412153 and by the Cottrell Scholar Award from the Research Corporation for Science Advancement.
Support for PFH was provided by an Alfred P. Sloan Research Fellowship, NASA ATP Grant NNX14AH35G, and NSF Collaborative Research Grant \#1411920 and CAREER grant \#1455342.
Numerical calculations were run on the Caltech compute cluster ``Zwicky'' (NSF MRI award \#PHY-0960291), through allocation TG-AST120025, TG-AST130039 and TG-AST150045 granted by the Extreme Science and Engineering Discovery Environment (XSEDE) supported by the NSF, and through NASA High-End Computing (HEC) allocation SMD-14-5189, SMD-14-5492, SMD-15-5950, and SMD-16-7592 provided by the NASA Advanced Supercomputing (NAS) Division at Ames Research Center.

\bibliographystyle{mnras}
\bibliography{deuterium}

\bsp

\label{lastpage}

\end{document}